\documentclass[preprint,5p,twocolumn]{elsarticle}

\usepackage{amsfonts}
\usepackage{amsthm}
\usepackage{amsmath}
\usepackage{amssymb}
\usepackage{color}

\def\z{\zeta}

\def\R{\mathbb{R}}
\def\ra{R}
\def\r{R}

\def\C{\mathbb{C}}

\def\x{\boldsymbol{x}}

\def\W{W}
\DeclareMathOperator*{\under}{\longrightarrow} 

\begin{document}

\begin{frontmatter}




\title{Do scale-invariant fluctuations imply the breaking of de Sitter invariance?}


\author{A. Youssef}\ead{youssef@mathematik.hu-berlin.de}

\address{Institut f\"ur Mathematik und Institut f\"ur Physik, Humboldt-Universit\"at zu Berlin\\
Johann von Neumann-Haus, Rudower Chaussee 25, 12489 Berlin, Germany}

\begin{abstract}
The quantization of the massless minimally coupled (mmc) scalar field in de Sitter spacetime is known to be a non-trivial problem due to the appearance of strong infrared (IR) effects. In particular, the scale-invariance of the CMB power-spectrum -  certainly one of the most successful predictions of modern cosmology - is widely believed to be inconsistent with a de Sitter invariant mmc two-point function. 
 Using a Cesaro-summability technique to properly define an otherwise divergent Fourier transform, we show in this Letter that de Sitter symmetry breaking is \emph{not} a necessary consequence of the scale-invariant fluctuation spectrum. We also generalize our result to the tachyonic scalar fields, i.e the discrete series of representations of the de Sitter group, that suffer from similar  strong IR effects.
\end{abstract}


\end{frontmatter}



\section{Introduction: Fourier versus coordinate-space two-point function}
\label{}
We review here the standard material leading to the prediction of a scale-invariant power-spectrum for the CMB fluctuations. Consider the mmc scalar field action:
$$
S=-\frac{1}{2}\int d^4x \sqrt{-g} \:g^{\mu \nu} \partial_\mu \phi \partial_\nu \phi, 
$$ 
where $g_{\mu \nu}$ is the de Sitter metric and $g$ its determinant. Making the change of variable $u=a \phi$, where $a$ is the scale factor, the quantum field can be written as 
\begin{equation*}
\label{FieldFourier}
\hat{u}(\tau,\x)=\frac{1}{(2\pi)^{3/2}}\int d\boldsymbol{k}\left[  \hat{a}_{\boldsymbol{k}} \:u_{k}(\tau) e^{i \boldsymbol{k}.\x}+ \hat{a}_{\boldsymbol{k}}^\dagger \:u_{k}^*(\tau) e^{i \boldsymbol{k}.\x}\right]
\end{equation*}
where $\tau$ is the conformal time defined below.  In the Bunch-Davies  vacuum state \cite{Birrell:1982ix}, the normalized mode functions $u_{k}$ read
$$
u_k= \sqrt{\frac{\hbar}{2k}} e^{-i k \tau} \left(1-\frac{i}{k\tau} \right).
$$
The power-spectrum $\mathcal{P}(k)$ is defined through the Fourier transform:
$$
\left<0\vert \phi(\x,\tau) \phi(\x',\tau)\vert0\right>=\int d\boldsymbol{k}\: e^{i \boldsymbol{k}. (\x-\x')} \frac{\mathcal{P}(k)}{4\pi k^3}.
$$
This gives
\begin{equation}
\label{pk}
\mathcal{P}(k)=\frac{\vert u_k \vert^2}{a^2} \frac{k^3}{2\pi^2}=\left(\frac{H}{2\pi}\right)^2 \left[1+ \frac{k^2}{a^2 H^2} \right]
\end{equation}
where $H$ is the Hubble constant. When the wavelength is much larger than the Hubble radius, one gets the celebrated scale-invariant power spectrum:
$$
\mathcal{P}(k) \approx \hbar \left(\frac{H}{2\pi}\right)^2.
$$

On the other hand, it is widely known that the construction of a coordinate-space representation of the mmc field two-point function in de Sitter has remained a matter of controversy and subject of debate for decades. 
Indeed, IR divergences arise in the quantization of the mmc field, leading to important technical and conceptual questions about the breakdown of de Sitter invariance \cite{Allen:1987tz,Wood} and/or of perturbation theory \cite{Hollands,Kitazawa,Raja}. 

We address in this Letter this tension between the Fourier and coordinate-space representation of the two-point function. As a direct and important consequence, our work supports the possibility of a de Sitter-invariant quantization of the mmc field that also agrees with the observed scale-invariant CMB power-spectrum.

The organization of this Letter is as follows: after exposing the necessary basics of de Sitter geometry and QFT, we review the appearance of IR divergences in the coordinate-space two-point function. We then present the construction given in \cite{MR2679970} to deal with these divergences. In section \ref{scaleinvsymbreaking} we expose the central contribution of this Letter, namely the use of a Cesaro-summation technique to define the otherwise divergent Fourier transform that relates the power-spectrum to the coordinate-space two-point function. Finally we show that this method is robust enough to compute the power spectrum for all the scalar ``tachyonic" fields in de Sitter.

\section{De Sitter geometry}
The $d$-dimensional de Sitter spacetime can be identified with the real one-sheeted hyperboloid in the $d+1$ Minkowski spacetime $M_{d+1}$\footnote{We will use the ``mostly minus'' metric throughout the article
$$
x^2=\eta_{\mu \nu} x^\mu x^\nu, \quad \eta_{\mu \nu}=\textrm{diag}(1,-1,\cdots,-1).
$$}:
$$
X_d=\left\lbrace x \in \R^{d+1}, x^2=-\ra^2\right\rbrace
$$
with $\ra>0$ being the de Sitter ``radius''.
This definition of the de Sitter manifold reveals the maximal symmetry of $X_d$ under the action of the de Sitter group $SO_0(1,d)$. We define for convenience the de Sitter invariant quantity 
\begin{eqnarray*}
\z(x,x')&=\dfrac{x.x'}{R^2}, \quad x,x' \in X_d.
\end{eqnarray*}
Because of the causality properties of the Bunch-Davies vacuum $\z$ will vary in the cut-plane
$$
\C_\Delta=\C \setminus \Delta, \quad \Delta= \{ \z \in \C: \z<-1\}.
$$
\paragraph{Planar coordinates}
This is the coordinate system the most relevant to cosmology. Here the spatial sections are $d-1$ planes. Only half of the de Sitter spacetime is covered by this coordinate system and it reads:
$$
x(t,\x) = \left\{ 
\begin{array}{rl} 
x^0= & \ra \sinh \frac{t}{\ra} +\frac{1}{2\ra}  \x^{2} \: e^{\frac{t}{\ra}} \\ 
x^j=  &  \x^j \: e^{\frac{t}{\ra}},  \quad \quad \quad \quad  \quad \quad \x \in \R^{d-1}\\
x^d= & \ra\cosh \frac{t}{\ra} -\frac{1}{2\ra} \x^{2} \: e^{\frac{t}{\ra} }.
\end{array} \right.
$$
The de Sitter metric and the invariant quantity $\z$ in this coordinate system are given by
\begin{eqnarray*}
ds^2&=&dt^2-e^{2t/\ra} d\x^{2}\\
\z(x,x')&=& \frac{e^{\frac{t+t'}{\ra}} }{2\ra^2}(\boldsymbol{x}-\boldsymbol{x}')^2 -\cosh\left( \frac{t-t'}{\ra}\right).
\end{eqnarray*}
In particular we have
$$
\z(\x,t;\x',t)= \frac{e^{2t/\ra} }{2\ra^2}r^2 -1.
$$
Finally it is convenient to define the conformal time 
$$\tau=\int \dfrac{dt}{a(t)}.$$.

\section{The massless field in de Sitter}
\label{}


The physical reason behind the appearance of strong IR effects in de Sitter can be simply understood: the rapid expansion of the spacetime dilate correlation patterns. After all this is the exact reason why a de Sitter inflationary phase in the early  universe solves many problems of the hot big-bang model. These IR effects are present at the interacting level for massive fields (see  \cite{Marolf:2010nz,Akhmedov}). They are even stronger for massless (and non-conformally invariant) fields - such as the mmc and the graviton - as they appear already at the tree-level. We review here these IR divergences in the mmc case. 

Recall that in the Bunch-Davies vacuum state the two-point function for a massive scalar field reads :
\begin{align*}
\W_m(x,x')= &\Gamma(-\sigma) \: \frac{\Gamma(\sigma+d-1)}{(4\pi)^{d/2}\r^{d-2}\Gamma(\frac{d}{2})} \\
&\;  {_{2}}F_{1}\left[-\sigma,\sigma+d-1,\frac{d}{2};\frac{1-\z}{2}\right]
\end{align*}
where $\sigma=-\frac{d}{2}+\sqrt{\frac{d^2}{4}-m^2 \r^2}$ and $ {_{2}}F_{1}$ is the hypergeometric function.  In the massless limit, $\sigma \to 0$, this expression diverges because of the pole of the first Gamma function and we have the small mass expansion:
$$
\W_m(x,x') \approx \frac{d}{4 \pi^{\frac{d+1}{2}}} \Gamma\left(\frac{d-1}{2}\right)\frac{1}{m^2 \r^d}+ \textrm{regular terms in $m$}.
$$ 
Note that in the flat space limit ($\r\to \infty$), this singular term is absent and the massless limit is smooth\footnote{\label{noncomm}This means that the flat space limit ($\r\to \infty$) and the massless limit  ($m\to 0$) do not commute. This is a physically important fact and might mean that even a small amount of curvature - like in today's universe - might have important consequences on massless fields.}. More precisely  we have (for $d>2$):
\begin{align*}
\W_m^\textrm{flat}(x,x') &= -\frac{i \pi}{(4\pi)^{d/2}}\left(\frac{m^2}{\mu^2}\right)^{\frac{d-2}{4}} H^{(2)}_{\frac{d}{2}-1}\left( 2 \sqrt{m^2 \mu^2}\right)\\
& \under_{m \to 0} \left(\frac{1}{2\sqrt{\pi}}\right)^d  \Gamma\left(\frac{d}{2}-1\right) \mu ^{2-d}
\end{align*}
where $\mu=\sqrt{(x-x')^2}$ is the invariant distance and $H^{(2)}$ is the Hankel function of the second kind. 

One of the first papers studying the mmc scalar field  in de Sitter is \cite{Allen:1987tz}, where the authors prove that a usual de Sitter-invariant Fock space quantization is impossible in this case. They then propose to trade the de Sitter SO$(1,d)$ invariance for a smaller one, say a SO$(d)$ invariance. 
Equivalently, it is a common belief among workers in the field that a scale-invariant power-spectrum leads necessarily to a breakdown of de Sitter invariance and that some physical quantities might thus become time-dependent.

 Several authors later proposed different treatments of the mmc field, among which \cite{MR2679970} is one of the most exhaustive. Here the divergent term is  subtracted, a ``renormalized'' two-point function is computed and it reads:
\begin{equation}
\label{rw}
\W(x,x')=\frac{1}{8\pi^2\r^2} \left[ \frac{1}{1+\z}-\ln (1+\z)\right]+\textrm{constant}.
\end{equation}
Hence this procedure allows for a de Sitter-invariant quantization. The draw-back  is that the two-point function no longer verifies the equation of motion $\Box \phi=0$, instead it verifies the anomalous equation
$$
\Box \phi=-\frac{\Gamma\left( \frac{d+1}{2}\right)}{2 \pi^{\frac{d+1}{2}}}.
$$ 
This simple renormalization procedure has been used implicitly in several earlier works. However, the major contribution of \cite{MR2679970} is proving that on a suitably chosen subspace of states $\mathcal{E}$, the equation of motion is effectively restored. This ``Physical'' space of states should be regarded the same way as we regard the one that appears in the quantization of gauge theory (for instance the space of transverse photons in QED). Moreover, the authors were able to show that the renormalized two-point function defines a positive kernel when restricted to $\mathcal{E}$, thus enabling a probabilistic interpretation of the theory.

\section{Scale-invariance of the power spectrum and de Sitter symmetry breaking}
\label{scaleinvsymbreaking}

As explained before, we define the power spectrum by the Fourier transform
$$
\W(\x,t;\x',t)=\int d\boldsymbol{k}\: e^{i \boldsymbol{k}. (\x-\x')} \frac{\mathcal{P}(k)}{4\pi k^3}.
$$
The two-point function obtained from the scale-invariant part of the power spectrum (we set $\hbar=1$ in all following formulas):
$$
\mathcal{P}(k) \approx  \left(\frac{H}{2 \pi}\right)^2
$$ 
is logarithmically divergent in the IR. Hence this Fourier transform is only formal. As we explained earlier, this divergence is commonly believed to induce de Sitter symmetry breaking\footnote{A similar issue appears in the QFT of a massless scalar field in two-dimensional Minkowski spacetime. In this model too, an interesting discussion arises on the general interplay between IR singularities and the occurrence of Lorentz symmetry breaking \cite{Strocchi}. This similarity might however be more of a mathematical than a physical nature.}. 

We now present the central contribution of this Letter, namely the calculation of the power-spectrum obtained from the de Sitter-invariant renormalized two-point function \eqref{rw}. The latter is given by (up to a constant term that we will show to be irrelevant)
$$
\W(x,x')=\frac{1}{8\pi^2\r^2} \left[ \frac{1}{1+\z}-\ln (1+\z)\right].
$$
In spatially flat coordinates this gives
$$
\W(\x,0;\x',0)=\frac{1}{4 \pi ^2 r^2}-\frac{H^2 \ln \left(H^2 r^2\right)}{8 \pi^2}
$$
The power spectrum is then formally given by
$$
\mathcal{P}(k)= \frac{1}{(2\pi)^3} \int e^{-i \boldsymbol{k}.(\x-\x')} 4\pi k^3  \W(r).
$$
For the $\dfrac{1}{r^2}$ part we get
\begin{align*}
\frac{k^2}{4 \pi^2}.
\end{align*}
However the power spectrum of the logarithmic part is given by 
\begin{align*}
-\frac{H^2 k^2}{4 \pi ^3}  \int_0^\infty dr \: \ln \left(H^2 r^2\right) \sin (k r)
\end{align*}
and is divergent. The integrand is however highly oscillatory and turns out to be Cesaro-summable\footnote{One can also regulate the IR divergence through:
$$
\int_0^\infty dr\:g(k,r)  \to \frac{d}{dk} \int_0^\infty dr\: \int^k dk\: g(k,r)
$$
Note that this method introduces another divergence near $r=0$ and one has to separate the integration domain into two regions, one near $0$ and the other near infinity. Cesaro-summability is a more physically sound and efficient option and we will use it throughout this Letter.}:
\paragraph{Cesaro summability \cite{tit}}
The integral $\int_0^\infty f(x) dx$ is $\alpha$ Cesaro summable and denoted $(C,\alpha)$, if the limit
$$
\lim_{\lambda \to \infty} \int_0^\lambda \left(1-\frac{x}{\lambda} \right)^\alpha f(x) dx
$$
exists and is finite. If an integral is $(C,\alpha)$ summable for some value of $\alpha$, then it is also $(C,\beta)$ summable for all $\beta>\alpha$, and the value of the resulting limit is unchanged.

Taking $\alpha=2$, the regularized integral can be computed in closed form and the limit is 
\begin{align*}
&\lim_{\lambda \to \infty} \int_0^\lambda \left(1-\frac{r}{\lambda} \right)^2 \left[ -\frac{4\pi}{k}  \:r \sin(k r) \frac{1}{8\pi^2} \ln \left( \frac{H^2 r^2}{2} \right)
\right]\\
&=\left(\frac{H}{2\pi}\right)^2.
\end{align*}
The final result, after restoring time dependence is
$$
\mathcal{P}(k)=\left(\frac{H}{2\pi}\right)^2 \left[1+ \frac{k^2}{a^2 H^2} \right]
$$
which is exactly the power spectrum one gets for the de Sitter mmc field \eqref{pk} if we take the formal Fourier representation from the beginning as explained in the first section. 

We end this section by two remarks. First, note that the Cesaro technique is a summability technique in the mathematical sense. In particular it does not modify any convergent integral. Instead, it only gives meaning to a certain class of divergent integrals, moreover without the introduction of any arbitrary cutoff that has to be eliminated afterwards (as in \cite{Rajaraman:2010zx} for instance).

Second, note that any two-point functions that differ by a constant are Cesaro-summable to the same power-spectrum. This fact is quite interesting as the renormalization procedure presented in \cite{MR2679970} only gives the two-point function up to a constant. In our construction, the power-spectrum, which is a physical observable, is thus indifferent to this arbitrary constant term, a fact that is not obvious a priori.

 Finally, the mmc two point-function \eqref{rw} is logarithmically growing for largely separated points, a rather unconventional fact. However, we have seen that this is exactly what is needed in order to reproduce the observed scale-invariant power-spectrum. In other words, at least in this situation, \emph{the IR growing of the two-point function is physical}. This is a quite important observation, since such IR growing terms are often encountered in de Sitter and their meaning is still ill-understood  (see \cite{AhmedIR1} and references therein).


\section{Discrete series power spectrum}
This renormalization procedure (subtraction of the $1/m^2$ divergence) presented for the mmc scalar field has been generalized in \cite{MR2679970} to the tachyonic fields\footnote{We prefer the denomination ``discrete series of representation" to the ``tachyon" one used in \cite{MR2679970} because
the flat space limit of these fields is the massless field and not negative mass squared fields. Hence referring to these fields as the discrete series representations is more accurate. We will use however the two denominations in what follows.} of negative mass squared:
$$
m^2=-n(n+d-1),\quad n \in \mathbb{N}.
$$  
It was also proven that this renormalization scheme gives rise to a perfectly well-defined free QFT in de Sitter. We find that the corresponding two-point functions, denoted by $\W_n$, have a growing large distance behavior given by
$$
\W_n\sim \z^n \ln \z.
$$

The Cesaro-summation method we have been using for the mmc field is sufficiently robust and enables us, after some lengthy calculations,  to compute the power-spectrum of all the tachyonic fields. In terms of the variable $x=\frac{k}{H a}$, we obtain
\begin{align*}
\mathcal{P}_n&=\left(\frac{H}{2\pi}\right)^2 \left[ x^2+\sum_{m=0}^{n} \frac{a_{n,m}}{x^{2m}}\right],  \\
a_{n,m}&= \frac{1}{\sqrt{\pi}} \dfrac{\Gamma(\frac{3}{2}+m)\Gamma(3+n+m)}{\Gamma(2+m)\Gamma(1-m+n)}.
\end{align*}
The sum in this formula can be explicitly evaluated and we get 
\begin{align*}
\mathcal{P}_n&=\left(\frac{H}{2\pi}\right)^2\frac{\pi \:x^3}{2}   \left[J^2_{n+\frac{3}{2}}(x)+Y^2_{n+\frac{3}{2}}(x)\right]
\end{align*}
where $J_\nu$ and $Y_\nu$ are the Bessel functions of the first and second kind respectively. 
This calculation is a first step towards an eventual observable effect of these tachyons through their influence on the CMB power-spectrum. It might also permit to rule out their existence\footnote{As will be explained elsewhere, the non scale-invariant behavior for small $x$ does not necessarily mean that these tachyonic theories are ruled out by observation.}. Our results on this will be presented elsewhere.
\newline

Finally, the possible generalization of this Cesaro-summability technique to the interacting theory is a quite interesting question and could constitute a first step towards a more ambitious IR renormalization program for massless interacting fields in de Sitter space. This issue is studied in \cite{IRrenorm}.

\section*{Acknowledgement}
We acknowledge fruitful discussions with D. Langlois, F. Nitti, and D. Steer. We also wish to thank E. Akhmedov and R. Woodard for their constructive comments on an earlier version of this Letter.

\end{document}